\documentclass[oldversion]{aa}  
\usepackage{graphicx}
\usepackage{txfonts}
\usepackage{longtable,rotate}
\usepackage{booktabs}
\usepackage{lscape}
\usepackage{tabularx}
\usepackage{colortbl, xcolor}

\newcommand{\gray}{\rowcolor[gray]{0.9}}

\begin{document}

\title{GRB~060605: multi-wavelength analysis of the first GRB observed using
integral field  spectroscopy\thanks{Based on observations collected at the
German-Spanish Calar Alto Observatory in Spain (Programme F06-3.5-055) and
at the European Southern Observatory, La Silla and Paranal, Chile (ESO
Programme 177.D-0591).}}

\author{
Patrizia~Ferrero\inst{1},
Sylvio~Klose\inst{1},
David~Alexander~Kann\inst{1},
Sandra~Savaglio\inst{2},
Steve~Schulze\inst{1},
Eliana~Palazzi\inst{3},
Elisabetta~Maiorano\inst{3,4},
Petra~B\"ohm\inst{5},
Dirk~Grupe\inst{6},
Samantha~R.~Oates\inst{7},
Sebasti\'{a}n~F.~S\'{a}nchez\inst{8},
Lorenzo Amati\inst{3},
Jochen~Greiner\inst{2},
Jens~Hjorth\inst{9},
Daniele~Malesani\inst{9},
Scott~D.~Barthelmy\inst{10},
Javier~Gorosabel\inst{11},
Nicola~Masetti\inst{3},
\and
Martin~M.~Roth\inst{5}
}

\offprints{P. Ferrero}

   \institute{Th\"uringer Landessternwarte Tautenburg, Sternwarte 5,
   D--07778 Tautenburg, Germany\\
   \email{ferrero@tls-tautenburg.de}   
\and   
   Max-Planck-Institut f\"ur Extraterrestrische Physik, 
   Giessenbachstrasse, D--85748 Garching, Germany
\and
   INAF -- Istituto di Astrofisica Spaziale e Fisica Cosmica di Bologna,
   via Gobetti 101, I--40129 Bologna, Italy
\and
   Dipartimento di Fisica, Universit\`{a} di Ferrara, via Saragat 1, I--44100 
   Ferrara, Italy
\and
   Astrophysikalisches Institut Potsdam, An der Sternwarte 16, D--14482 
   Potsdam, Germany
\and
   Department of Astrophysics and Astronomy, Pennsylvania State University, 
   525 Davey Lab. University Park, PA 16802, U.S.A.
\and
   Mullard Space Science Laboratory, University College London, Holmbury St 
   Mary, Dorking, Surrey RH5 6NT, U.K.
\and
   Centro Astron\'{o}mico Hispano Alem\'{a}n de Calar Alto, 
   Calle Jesus Durban Remon 2-2, E--04004 Almer\'{\i}a, Spain
\and
   Dark Cosmology Centre, Niels Bohr Institute, University of Copenhagen, 
    Juliane Maries Vej 30, DK--2100 Copenhagen, Denmark
\and
   Astroparticle Physics Laboratory, Mail Code 661, NASA Goddard Space 
   Flight Center, Greenbelt, MD 20771, U.S.A.
\and
   Instituto de Astrof\'{\i}sica de Andaluc\'{\i}a (IAA-CSIC), Apartado 
   de Correos 3.004, E--18080 Granada, Spain
   }
   
\date{Received: 14 April 2008 / Accepted: 12 February 2009}
 
\authorrunning{Ferrero et al.}
\titlerunning{GRB 060605}

\abstract {
The long and relatively faint  gamma-ray burst GRB 060605 detected by
\emph{Swift}/BAT lasted about 20 sec. Its afterglow could be
observed with \emph{Swift}/XRT for nearly 1 day, while \emph{Swift}/UVOT could
detect the afterglow during the first 6 hours after the event. Here,  we
report on integral field spectroscopy of its afterglow performed with
PMAS/PPak mounted at the Calar Alto 3.5 m telescope. In addition, we report on
a detailed analysis of XRT and UVOT data and on the results of deep late-time
VLT observations that reveal the GRB host galaxy.
We find that the burst occurred at a redshift of $z$=3.773,  possibly
associated with a faint, $R_C=26.4 \pm 0.3$ host. Based on the optical and
X-ray data, we deduce information on the SED of the afterglow, the position of
the cooling frequency in the SED, the nature of  the circumburst  environment,
its collimation factor, and its energetics. We find that the GRB fireball was
expanding into a constant-density medium and that the explosion was collimated
with a narrow half-opening angle of about 2.4 degrees. The initial Lorentz
factor of the fireball was about 250; however, its  beaming-corrected energy release in
the gamma-ray band was comparably low. The optical, 
X-ray afterglow, on the other hand, was rather luminous.  Finally, we find
that the data are consistent within the error bars with an achromatic evolution
of the afterglow during the suspected  jet break time at about 0.27 days after
the burst.}

\keywords{Gamma rays: bursts: individual: GRB 060605}
\maketitle


\section{Introduction}

Since its launch in November 2004, the \emph{Swift} satellite (Gehrels et
al.~\cite{Gehrels2004}) has localised more than 300  gamma-ray
bursts (GRBs)  with an accuracy of 3 to 4 arcmin radius for the 
satellite-based analysis and 1 arcmin for the ground-based one, using the BAT detector
(Barthelmy et al. \cite{Barthelmy2005}). Of these, $\sim$84\%
could be localised with  the \emph{Swift} X-ray telescope (XRT; Burrows et
al.~\cite{Burrows2005}) and $\sim$72\% had an optical/near-infrared
afterglow. For $\sim$30\% of the entire sample, it was possible to measure a
redshift (see also J. Greiner's internet page at
\texttt{http://www.mpe.mpg.de/$\sim$jcg/grbgen.html}).

The most widely accepted GRB model is the fireball model (e.g., Cavallo \&
Rees~\cite{Cavallo1978}; Rees \& M\'{e}sz\'{a}ros~\cite{Rees1992}; Sari,
Piran, \& Narayan~\cite{Sari1998}; for reviews: Piran \cite{Piran2005};
M\'{e}sz\'{a}ros \cite{Meszaros2006}). Within its framework  the burst is
accompanied by a relativistic, collimated  outflow that sweeps up the
surrounding interstellar medium. The shocks that form convert the kinetic energy
of the flow into internal energy of accelerated particles, which in turn emit
synchrotron radiation from X-ray to radio wavelengths. The afterglows are
usually identified as either new objects in comparison to archival images  or
by their fading behaviour.  The study and detection of the afterglows enable
sub-arcsecond localisation  of the burst and unambigous determination of its
host galaxy and its redshift if the afterglow is bright enough.
The afterglow itself provides information about the physical processes that
work and can reveal clues to the nature of the central engine and to the
environmental properties of the progenitors. Most of these data can only be derived
via a spectroscopic analysis of the optical and X-ray afterglow.
 
Even though optical afterglows can be very bright at the beginning, the rapid
fading of these transients makes the timing of observations crucial for the
acquisition of spectroscopic data with a sufficient signal-to-noise (S/N)
ratio. However, due to the time usually needed to identify the optical
transient in a GRB X-ray error circle, rapid spectroscopic follow-up
observations are a challenge. Indeed long-slit spectroscopy has to await the
identification of the afterglow, or a best guess has to be made; i.e., if the
error box is very small, one can assume that  the afterglow is the brightest
object in the field. Integral field spectroscopy (IFS), on the other hand,
using integral field units (IFUs), can start getting spectra of an entire
error box as soon as an arcsecond X-ray location has been reported, usually in
the case of the \emph{Swift} satellite  within minutes  after the GRB
trigger. In principle, once the afterglow has been identified by other means,
IFS data could then be used to extract early spectra. This procedure would
minimise an important bias,  namely the pre-selection of afterglows for
spectroscopic follow-up observations according to their apparent magnitude at
the time of their discovery.  Furthermore, in the \emph{Swift} era, many
optical afterglows are discovered first by  the \emph{Swift} UV/optical
telescope which has only filters up to the $v$ band precluding the rapid
localisation of $z\gtrsim5$ or of highly extinguished afterglows (cf. Roming
et al. \cite{Roming2006}).  Needing only \emph{Swift} XRT localisations, IFS
is basically  not affected by this colour-selection bias.

Motivated by the aforementioned potential advantages of IFS, we have started
an observing campaign of GRB afterglows with IFUs. Here, we report on our
first successful IFS observations of an afterglow (GRB 060605) performed at
the Calar Alto  3.5 m telescope. In addition, we report on the analysis of the
\emph{Swift} BAT, XRT and UVOT data and late-time VLT observations.

Throughout this paper we adopt a world model with $H_0=71$ km s$^{-1}$
Mpc$^{-1}, \Omega_{\rm M}=0.27, \Omega_\Lambda=0.73$ (Spergel et
al. \cite{Spergel2003}). For the flux density of the afterglow we use 
the usual convention $F_\nu(t) \propto t^{-\alpha} \nu^{-\beta}$.

\section{Observations and data reduction}

\subsection{Swift BAT data: the burst}

GRB 060605 was detected by the BAT instrument on-board \emph{Swift} on June 5,
at $T_0$= 18:15:44.61 UT (trigger 213630; Page et al.~\cite{Page2006}) with an
accuracy of 3 arcmin radius (90\% containment, including systematic
uncertainty). The BAT on-board calculated location of the burst was
R.A. (J2000) = 21$^h$ 28$^m$ 35$^s$ and Decl. = --06$^{\circ}$ 3\arcmin
36\arcsec (Page et al.~\cite{Page2006}), while ground analysis resulted in
coordinates R.A. (J2000) = 21$^h$ 28$^m$ 37.6$^s$ and Decl. = --06$^{\circ}$
2\arcmin 44\farcs7 with an accuracy of 1.5 arcmin radius.

The time-averaged spectrum of the burst (from $T_0 - 2.580$ s to $T_0 +
20.450$ s) can be described by a cutoff power law with  $\alpha
=0.3_{-0.9}^{+0.7}$, and the peak energy at $90_{-20}^{+150}$ keV (Butler et
al. \cite{Butler2007}).  According to Sato et al. (\cite{Sato2006}), in the
15-350 keV band the burst had a duration of $T_{\rm 90} = 15\pm2$ s, while
according to  Butler et al. (\cite{Butler2007}) $T_{\rm 90} = 19\pm1$ s.

\begin{figure}
\centering
\includegraphics[width=8.5cm]{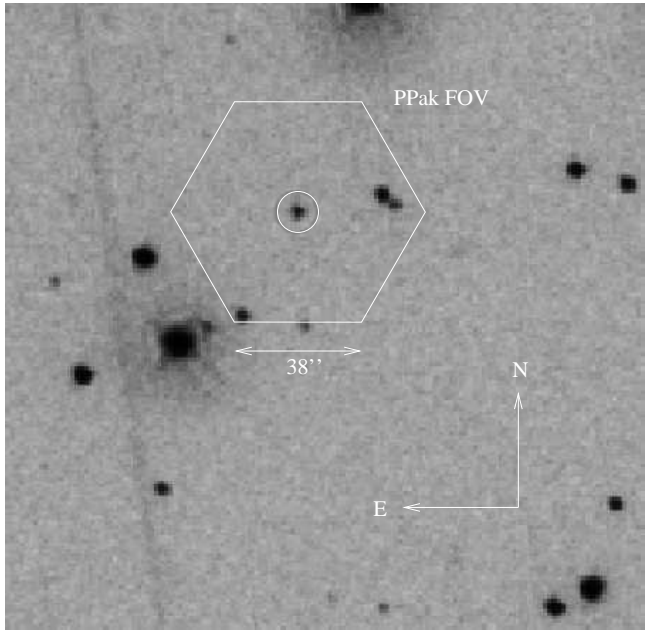}
\caption{
\emph{Swift} UVOT $v$-band image of the field of GRB 060605. The optical
afterglow is indicated by a circle. The overplotted hexagon shows the sky
coverage of  PMAS/PPak during our observing campaign (see also
Fig.~\ref{fig:spaxel}).}
\label{fig:field}
\end{figure}

\begin{figure}
\centering
\includegraphics[width=7.cm,angle=180]{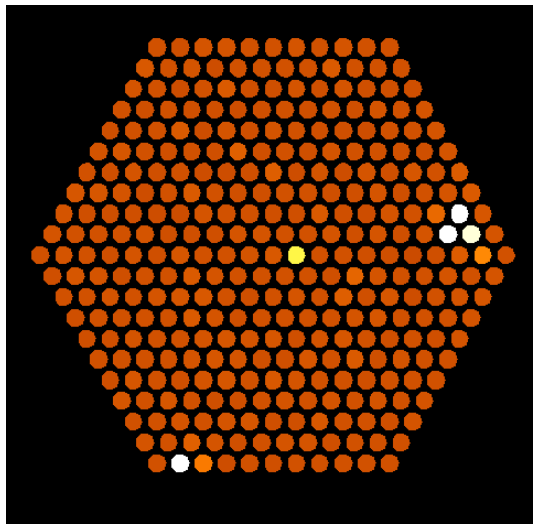}
\caption{The field of GRB 060605 seen by the PPak spaxels.  The image is the 
average of  three exposures when the afterglow (the yellow spot in the 
middle) had a magnitude of about $R_C$=19.5. The other bright spaxels
are stars (see Fig.~\ref{fig:field}).}
\label{fig:spaxel} 
\end{figure}

\subsection{Swift XRT data}

X-ray data of the afterglow of GRB 060605 were collected on 5 and 6 June 2006
with XRT.  Pointed observations on target started 93 s after the BAT  trigger
and the monitoring was organized in two sequences, with a total net exposure
time of $\sim$ 37.4 ks in photon counting (PC) mode and $\sim$ 13 s  in
windowed timing (WT) mode. In order to obtain a better  S/N ratio in the
spectral analysis, only the data of the first $\sim$ 30.5  ks of the  PC mode
observation were used.

The data reduction was performed using the XRTDAS v2.0.1 standard data 
pipeline package (\texttt{xrtpipeline} v0.10.3), in order to produce the final 
cleaned event files.

During sequence 000 the count rate of the burst was high enough to cause
pile-up in the PC mode data, which covered the entire first three orbits of
XRT observation from $T_{0}$ + 126 s to about $T_{0}$ + $1.8 \times  10^{4}$
s. Therefore, to account for this effect, the PC data were extracted  in a
circle of 25 pixels radius, with a circular region of 4 pixels radius excluded
from its centre. The size of the inner region was determined  following the
procedure described in Vaughan et al. (\cite{Vaughan2006}).

The X-ray background was measured within a circle with 40 pixels  radius
located far from any source. The ancillary response file was  generated with
the task \texttt{xrtmkarf} (v0.5.2) within
FTOOLS\footnote{\texttt{http://heasarc.gsfc.nasa.gov/ftools/}} (Blackburn
\cite{Blackburn1995}), and accounts for the size of the extraction region.  We
used the latest spectral  redistribution matrices
(swxpc0to12\_20010101v008.rmf) in the Calibration
Database\footnote{\texttt{http://heasarc.gsfc.nasa.gov/docs/heasarc/caldb/\\
caldb\_intro.html}} (CALDB 2.3) maintained by HEASARC.

\subsection{Swift UVOT data}

\emph{Swift} started settled observations of GRB 060605 with its UV/Optical
Telescope  (Roming et al. \cite{Roming2005}) 78 s  after the trigger.  The
very  first image was in the $v$-band, while the satellite was slewing.
\emph{Swift}  found an afterglow at coordinates R.A. (J2000) = 21$^h$ 28$^m$
37\fs32  and Decl. = --06$^{\circ}$ 3\arcmin 31\farcs3 (Page et
al.~\cite{Page2006}), confirming  the optical transient already identified at
that time by the robotic ROTSE IIIa telescope (Rykoff et
al.~\cite{Rykoff2006}).

The afterglow was only detected in the $white$, $v$ (see
Fig.~\ref{fig:field}),  and $b$ filters.  The lack of detection in the UV
filters (Blustin \& Page~\cite{Blustin2006}) is consistent with the redshift
of $z=3.7-3.8$ based on observations with the Australian National University
ANU 2.3-m  (Peterson \& Schmidt~\cite{Peterson2006}) and the  10-m Southern
African Large Telescope (SALT; Still et al.~\cite{Still2006}).

The initial observations, namely the $white$ and $v$ finding charts,
were  performed in event mode (photon counting), while the rest of the
exposures were taken predominately in image mode.

The source counts were extracted using a region of 5$\arcsec$ radius. As the
source  fades it is more accurate to use smaller source apertures (Poole et
al. \cite{Poole2008}). Therefore,  when the count rate fell below 0.5 counts
s$^{-1}$, the source counts were  extracted using a region with 3$\arcsec$
radius. These counts were corrected to 5$\arcsec$ using  the curve of growth
contained in the calibration files. Background counts were  extracted using a
circular region of radius 15$\arcsec$ from a blank area of  sky situated near
to the source position. The count rates were obtained from the event lists
using \texttt{uvotevtlc} and from the images  using \texttt{uvotsource}.  The
used software can be found in the software release, Headas 6.3.2 and  version
20071106 (UVOT)  of the calibration files.

For each filter, the count rates were binned by taking the weighted average in
time bins of $\Delta t/ t = 0.2$. They were then converted to magnitudes using
the UVOT photometric zero points (Poole et al.~\cite{Poole2008}).

\subsection{Spectroscopic data}

Low-resolution integral field spectroscopy of the field  was acquired starting
about 7.5 hours after the burst. Even if at that time the afterglow position
was already precisely known, we decided to perform the IFS observing run,
in order to learn the handling of the data.

The observations were carried out  starting at UT 01:43:41 (June 6), at the
3.5-m telescope equipped with the Potsdam Multi-Aperture Spectrograph (PMAS;
Roth et al.~\cite{Roth2005}) in the PPak (PMAS fiber Package) mode (Verheijen
et al.~\cite{Verheijen2004}; Kelz et al.~\cite{Kelz2006}), using  2$\times$2
pixel binning.  We used the V300 grating, which covers a wavelength range
between 3698 and 7010~\AA, resulting in a reciprocal dispersion of 3.4~\AA\
per pixel.  The PPak fiber bundle consists of 382 fibers of 2\farcs7  diameter
each (see fig. 5 in Kelz et al.~\cite{Kelz2006}). Of them, 331 fibers (the
science fibers) are concentrated in a single hexagonal bundle covering a
field-of-view of 74$''$ $\times$ 64$''$ with a filling factor of $\sim$65\%.

The sky is sampled by 36 additional fibers, distributed in 6 bundles of 6
fibers each, located following a circular distribution at $\sim$90$''$ from
the center and at the edges of the central hexagon. The sky-fibers are
distributed among the science ones in the pseudo-slit, in order to have a good
sampling of the sky. The remaining 15 fibers are used for calibration purposes.

Given that PMAS/PPak  has a filling factor less  than 1, during the
observations a dithering scheme was applied. The observations consisted of 9
single  exposures of 15 min each, i.e. 3 images for every dither pointing.  As
3 of the 9 exposures had a low S/N due to the presence of clouds, only six of
them were considered. Figure~\ref{fig:spaxel} shows  the average of the second
dither pointing images, when the magnitude of the afterglow was about
$R_C$=19.5.  The field of view is sampled into discrete spatial elements named
SPAXELs.

The data reduction was performed twice using two different pieces of software:
PPAK\_online,  which is part of the P3D package of IDL routines developed for
the reduction of PMAS data (Becker~\cite{Becker2002}) and
R3D\footnote{\texttt{http://www.caha.es/sanchez/r3d/index.html}}, a package
coded in Perl by S. F. S\'{a}nchez  (S\'{a}nchez \& Cardiel
\cite{Sanchez2005}; S\'{a}nchez \cite{Sanchez2006}).  In combination with the
previous ones, IRAF\footnote{\texttt{http://iraf.noao.edu}},
MIDAS\footnote{\texttt{http://www.eso.org/projects/esomidas}}  and the
E3D\footnote{\texttt{http://www.aip.de/Euro3D/E3D/\#Docu}} visualization tool
(S\'{a}nchez~\cite{Sanchez2004}) were used.  The results obtained using
the two packages were consistent.

The reduction of spectroscopic data obtained with fiber-based integral-field
units consists of the following standard steps: bias subtraction, flat field
correction, location of the spectra on the CCD (the so-called tracing), spectra
extraction, wavelength calibration, fiber flat correction, sky subtraction, 
cosmic ray rejection and flux
calibration.

The bias frame, obtained immediately after the target frame, was cleaned and
smoothed using boxsizes of 5 pixels in $x$ and $y$ to create the final bias
frame.  Domeflat exposures of 5 s were taken before and after the object
observations to produce a trace mask, i.e. to locate the spectra along the
cross-dispersion direction on the CCD (for a detailed description of tracing,
see Becker~\cite{Becker2002}).  Once this mask is defined one can easily
extract the spectra from the CCD, producing a so-called row-stacked-spectra
image, where one row represents one spectrum.

For wavelength calibration a combined He/Rb-emission lamp exposure of 15 s was
obtained at the beginning of the night with the additional illumination of 15
separate calibration fibers with ThAr. Simultaneous ThAr-exposures of these
calibration fibers  during lamp flat and object observations as well were used
to correct for flexure effects of the instrument (Kelz et
al.~\cite{Kelz2006}). We defined some (at least two) of the ThAr spots in the
lamp flat image as reference and calculated their shifts in $x$ and $y$ versus
the same ThAr spots in the object images.  These shift values were taken into
account during tracing, spectra extraction, and wavelength calibration as well.

For the sky subtraction the spaxels not contaminated by sources were selected
and the average extracted spectrum was then subtracted from the science
spectrum.  For this purpose we used the E3D package (S\'{a}nchez 
\cite{Sanchez2004}).

After cosmic ray rejection, the final spectrum was flux calibrated using the
spectrophotometric standard star Hz~44 (Oke~\cite{Oke1990}).  As the spectra of
the optical afterglow were extracted on one spaxel, the  spectra of the
standard star were extracted in the same way.  A cross check on the flux
calibration was performed using the observed $R_C$-band photometric magnitude.

\section{Results}

We first present here the results of our spectroscopy since this provided 
the accurate redshift information.

\subsection{The optical spectrum of the afterglow \label{NHoptical} }

\begin{figure}
\centering
\includegraphics[width=8.5cm]{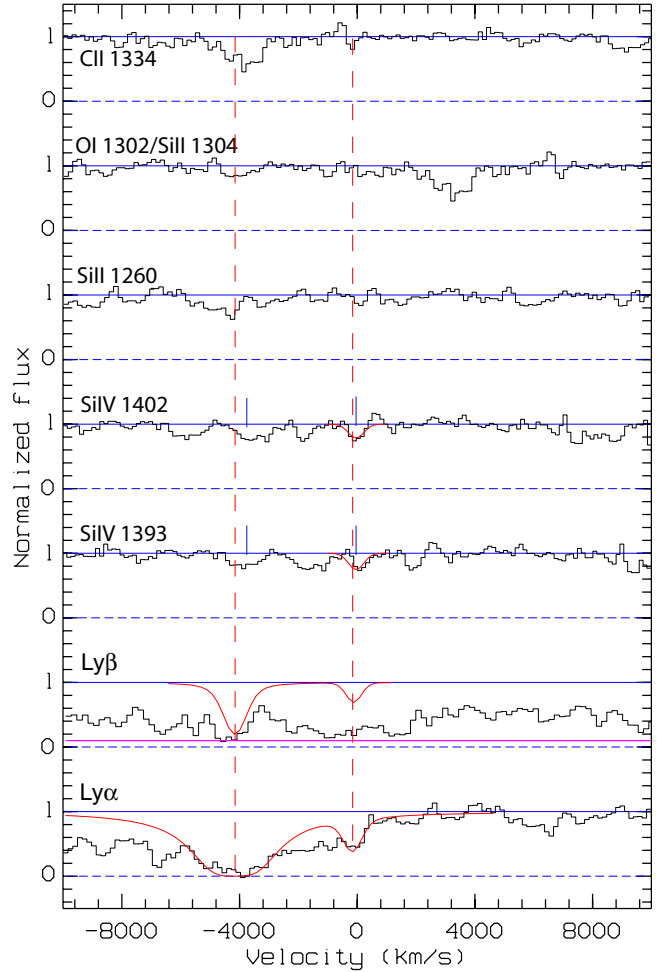}
\caption{The spectrum of the afterglow of GRB 060605 in velocity space,
centered at a redshift of $z$ = 3.773 ($v = 0$ km s$^{-1}$).  This is likely
the redshift of the GRB, for which we detect Ly~$\alpha$ and \ion{Si}{IV}
doublet absorptions. Offset by about $-4000$ km s$^{-1}$  we also mark the
strong absorption system (damped Lyman~$\alpha$) at $z$ = 3.709 with
Ly~$\alpha$, Ly~$\beta$, \ion{O}{I} and the \ion{Si}{IV} doublet, likely
associated with an intervening galaxy. For the method see Savaglio \& Fall
(\cite{Savaglio2004}).}
\label{fig:doublez}
\end{figure}

In Fig.~\ref{fig:doublez} we show absorption lines identified in  the PPak
spectrum of the afterglow (spectral resolution  $\lambda /\Delta \lambda$ =
500 at a wavelength of 5300 \AA). The highest redshifted Ly~$\alpha$ is at
$z$= 3.773 $\pm$ 0.001, which we interpret as the redshift of the GRB
(look-back time 11.98 Gyr). The \ion{H}{I} column density is very uncertain,
in the range $N_{\rm HI}$ = 10$^{18.5}$-10$^{19.3}$ cm$^{-2}$, but certainly
one of  the lowest ever measured in a GRB afterglow at the redshift of the GRB
(Savaglio \cite{Savaglio2006}; Jakobsson et al. \cite{Jakobsson2006}; Chen et
al.  \cite{Chen2007b}). We notice that the \ion{H}{I} column densities
measured for  GRB~021004 and GRB~060607A are also low, $N_{\rm HI}$ =
10$^{19.5}$ cm$^{-2}$ and 10$^{16.8}$ cm$^{-2}$, in the former and latter,
respectively. For GRB~030226, Shin et al. (\cite{Shin2006}) report $N_{\rm
HI}$ = 10$^{20.5 \pm 0.3}$ cm$^{-2}$. Possible explanations for 
such low \ion{H}{I} column densities might be 
either the ionization of the GRB environment
by the intense fireball light, the localisation of the burst source in a
star-forming region including many UV-bright massive stars, or  
the location of the GRB progenitor in the outer part of its host galaxy.

Blueward of the $z$ = 3.773 we identify a strong Ly~$\alpha$ absorber at  $z$
= 3.709 $\pm$ 0.003 ($\Delta v$ = 4000  km s$^{-1}$ from the GRB redshift)
likely associated with a Damped Ly~$\alpha$ system (DLA), with an estimated
\ion{H}{I} column density of $N_{\rm HI}$ = 10$^{20.9}$ cm$^{-2}$.  Redward of
the Ly~$\alpha$, we detect the Si~IV doublet at $z$ = 3.717 $\pm$ 0.001
($\Delta v$ = 500 km s$^{-1}$ from the DLA).  At approximately the redshift of
the DLA, we identify absorption lines  associated with \ion{C}{II} 1334,
\ion{Si}{II} 1260, and  \ion{O}{I} 1302/ \ion{Si}{II} 1304
(Fig.~\ref{fig:doublez}). Unfortunately, the low S/N of the
spectrum does not allow us to measure column densities for metals. We also
identify a strong \ion{Si}{IV} absorption doublet at $z$ = 3.774 $\pm$ 0.001,
$\Delta v$ = 120 km s$^{-1}$ redward of the Ly~$\alpha$, likely associated
with the GRB-host system. 

\begin{figure*}[!h,t]
\centering
\includegraphics[width=16.5cm]{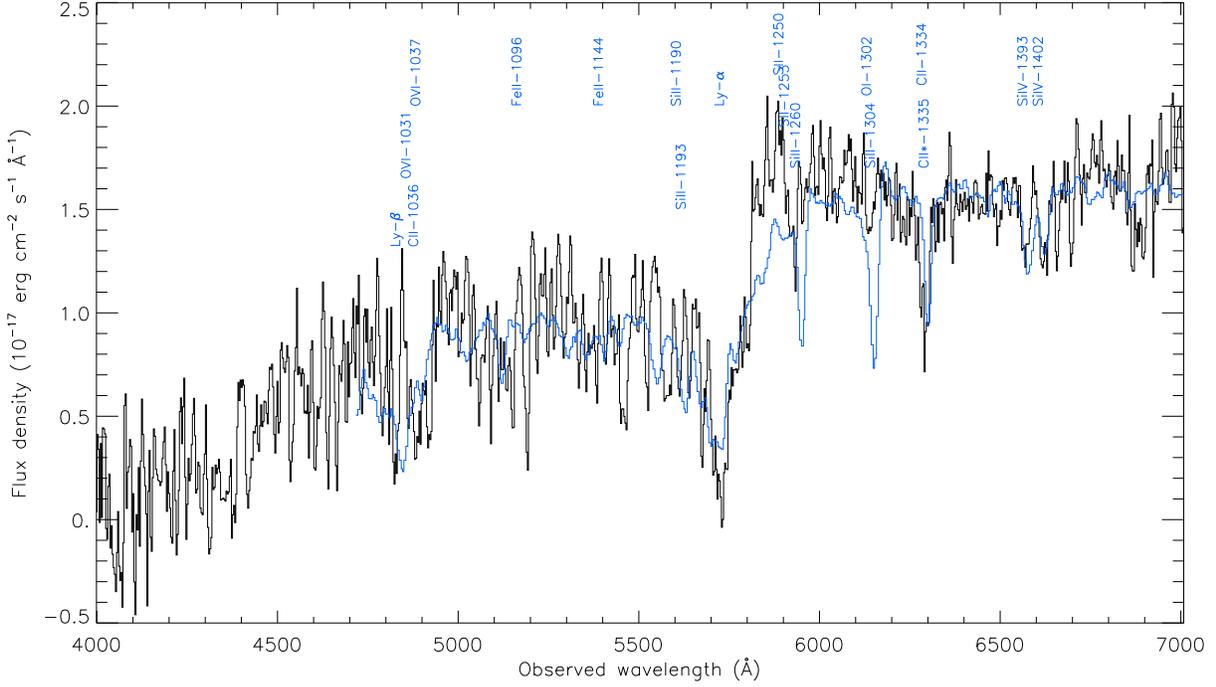}
\caption{Spectrum of the afterglow (black) of GRB 060605 obtained  between
7.5 and 9.1 hours after the burst with PPak mounted at the Calar Alto 3.5-m
telescope, overplotted with the spectrum of a Lyman break galaxy (LBG) at the
same redshift (from Shapley et al. \cite{Shapley2001}). The typical LBG lines
are indicated in blue; only the strongest of them can also be found in the 
afterglow spectrum. The afterglow spectrum, calibrated in flux and wavelength,
is a composition of six individual spectra of 15 min exposure time each.
The spectral resolution is $\lambda /\Delta \lambda$ = 500.}
\label{fig:LBGonGRB}
\end{figure*}

The separation between the two strong $z$ = 3.717 and $z$ = 3.774  Si~IV
absorbers $\Delta v$ = 3600 km s$^{-1}$ is comparable to that between  the
double \ion{C}{IV} absorbers detected in the afterglows of GRB 021004 ($\Delta
v$ = 2400 km s$^{-1}$; Savaglio et al. \cite{Savaglio2002}; Fiore et
al. \cite{Fiore2005}), GRB 030226  ($\Delta v$ = 2400 km s$^{-1}$; Klose et
al. \cite{Klose2004}), and GRB 060607A ($\Delta v$ = 1800 km s$^{-1}$;
Smette et al., in preparation). The possibility that this is the
signature of the stellar wind from the GRB progenitor (a Wolf-Rayet star) 
has been discussed in the literature in detail (Mirabal et al.~\cite{Mirabal2003};
Schaefer et al.~\cite{Schaefer2003}) but recently disfavoured for most cases by
Chen et al.~(\cite{Chen2007a}). The latter authors suggested as a reason for
this effect the presence of a foreground galaxies along the sight line of the
GRBs. Our deep VLT imaging of the field (Sect.~\ref{host})
does unfortunately not identify these potential absorbers with certainty. 
So, which of these two possibilities explains our observations 
of GRB 060605 remains open.

In Fig.~\ref{fig:LBGonGRB}, the spectrum of a Lyman break galaxy (LBG) at a $z
\sim 3$ (from Shapley et al. \cite{Shapley2001}) is overplotted  on the
spectrum of the GRB to show the typical absorption lines observed in these
galaxies at high redshifts and to help the comparison with our spectrum.  The
spectrum of the galaxy has been shifted in wavelength considering the redshift
of the damped Lyman~$\alpha$ absorption system at $z$ = 3.709 and rescaled in
flux for comparison. While the data seem to indicate an underabundance of
oxygen in the GRB host galaxy, the low S/N ratio of our data does
not allow us to draw quantitative conclusions.

\subsection{The X-ray afterglow \label{xray} }

\begin{figure}
\centering
\includegraphics[width=8.0cm]{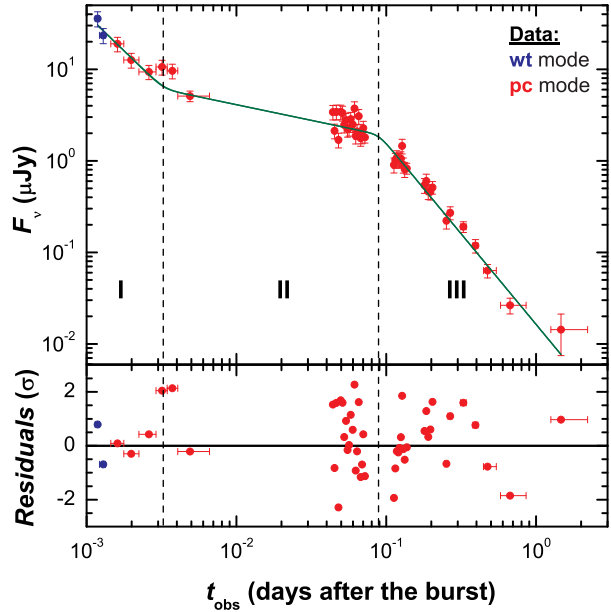}
\caption{
The X-ray light curve of the afterglow of GRB 060605 observed with
\emph{Swift} XRT (Evans et al.~\cite{Evans2007}).  Formally, three power-law
segments can be distinguished (Godet et al. 2006). A small flare is seen at the
beginning of the second decay phase (II). The lower panel shows the residuals
of the best fit. Fluctuations in the light curve ($>2\sigma$) are also seen
at later times. The energy conversion factor is 
4.6$\times$10$^{-11}$ erg cm$^{-2}$ counts$^{-1}$.}
\label{fig:lc}
\end{figure}

The X-ray afterglow of GRB 060605 was detected by \emph{Swift} for more than 1
day after the trigger. When analyzing the data  we rebinned them by taking 30
counts/bin in order to obtain a good S/N ratio. As already noted by Godet et
al. (2006), and as it is shown in (Fig.~\ref{fig:lc}), as a first guess the 
X-ray light curve consists only of three power-law segments. 
The numerical ansatz to describe the X-ray light curve is then a
smoothly broken \it double \rm power-law (see Liang et al. \cite{Liang2008}).
In doing so, we fixed the smoothness parameters $n_1$ and $n_2$ (in their paper
$\omega_1$ and $\omega_2$) to $-$10 and 10, in the case of transition I to
II and transition II to III, respectively. This describes a sharp break. For
the steep-to-shallow transition (I to II), we find a break time of $210\pm30$
s ($0.0024 \pm 0.0003$ days), while the shallow-to-steep transition (II to
III) took place at $7510\pm410$ s ($0.0869 \pm 0.0047$ days) after the
trigger. At the beginning, the afterglow decays with a slope of $\alpha_{\rm
I}=2.19\pm0.42$, followed by $\alpha_{\rm II}=0.34\pm0.03$ during the shallow
decay phase, and it continues to decay with $\alpha_{\rm III}=1.89\pm0.07$
($\chi^2/{\rm d.o.f} = 55.5/42=1.32$). Within errors, these values are in
agreement with the results reported by Godet et al. (\cite{Godet2006}). In
Sect.~\ref{Xcurve} we will test if this ansatz of a three-segment X-ray light
curve is compatible with the basic theoretical concepts describing afterglows.

\begin{figure}
\centering
\includegraphics[angle=270,width=8.5cm]{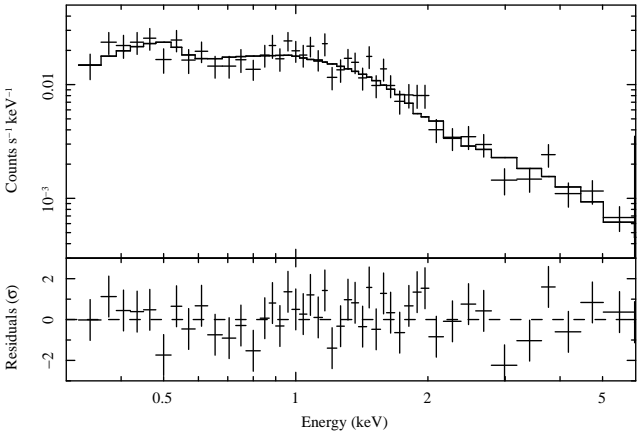}
\caption{X-ray spectrum of the afterglow of GRB 060605 obtained in photon
counting mode between  0.0015 and 0.8565 days after the trigger.
The lower panel shows the residuals of the best fit (for more 
details see Sect. \ref{xray}).}
\label{fig:sed}
\end{figure}

After dividing the 0.3-6 keV\footnote{The 0.3-10 keV XRT spectrum had no
signal in the range  6-10 keV and for this reason only the first part was
considered.}  XRT spectrum, derived from PC data, in several spectra over
small time intervals,  as no spectral evolution  was found, we took the
overall spectrum between $t=126$ s (0.0015 days) and $t=7.4\times10^4$ s (0.86
days) (Fig.~\ref{fig:sed}). It is well fitted  by an absorbed power-law with a
spectral index $\beta_{\rm X} = 1.06\pm0.16$ ($\chi^2/$d.o.f.=39.6/44=0.93)
and $N_{\rm H}$=5.5$^{+3.3}_{-2.9}\,\times\,10^{20}$ cm$^{-2}$. Due to the big
uncertainty this last value is consistent with both the Galactic hydrogen
column density of $N_{\rm H}^{\rm Gal}$=5.1$\,\times\,10^{20}$ cm$^{-2}$ by
Dickey \& Lockman (\cite{Dickey1990}) and the lower value of $N_{\rm H}^{\rm
Gal} = 4.1\,\times\,10^{20}$ cm$^{-2}$ given by the recent release of the
Leiden/Argentine/Bonn (LAB) Survey of Galactic \ion{H}{I} (Kalberla et
al. \cite{Kalberla2005}). No additional rest frame hydrogen column density can
be found in the X-ray spectrum. Adding such a component by hand did not
improve the fit. The lack of evidence of additional hydrogen in the host
galaxy is in agreement with the finding of Grupe et al. (\cite{Grupe2007a})
that high-redshift events usually do not show such a feature.

However, due to the large uncertainty on $N_{\rm H}$ (Ly~$\alpha$) and no
strong constraints on $N_{\rm H}$ (X-ray), we cannot exclude that for GRB
060605 the optical and the X-ray data trace a different population  of
hydrogen at the redshift of the burst, as it has been found in many other
cases (Watson et al. \cite{Watson2007}).


\subsection{The UV/optical light curve and its SED \label{optcurve}}

\begin{table}
\begin{minipage}[t]{\columnwidth}
\caption{Log of the \emph{Swift} UVOT observations.}
\centering
\renewcommand{\footnoterule}{}
 \begin{tabular}{rrccl}
\toprule
$t$	&$\Delta t -$	&$\Delta t +$	&	Magnitude &	Filter	\\
\midrule
\gray 106.6&4.9&5.1& $18.23_{-0.16}^{+0.18} $ &$white$\\
\gray 126.6&4.9&5.1& $17.97_{-0.13}^{+0.15} $ &$white$\\
\gray 151.6&7.3&7.7& $18.06_{-0.12}^{+0.13} $ &$white$\\
\gray 181.5&7.3&7.6& $17.83_{-0.10}^{+0.11} $ &$white$\\
\gray 4690&49&49& $19.81_{-0.14}^{+0.16} $ &$white$\\
\gray 6122&49&49& $20.14_{-0.24}^{+0.31} $ &$white$\\
\gray 17560&193&195& $20.90_{-0.23}^{+  0.30} $ &$white$\\
85.7&7.6&8.4& $17.72\pm0.66$ &$v$\\
220&9&9& $16.66_{-0.14}^{+0.16} $ &$v$\\
263&12&13& $16.69_{-0.12}^{+0.14} $ &$v$\\
318&15&15& $16.37_{-0.09}^{+0.10} $ &$v$\\
383&17&18& $16.67_{-0.11}^{+0.12} $ &$v$\\
463&22&23& $16.45_{-0.09}^{+0.10} $ &$v$\\
551&21&22& $16.60_{-0.10}^{+0.12} $ &   $v$\\
5100&49&49&$18.45_{-0.18}^{+0.22} $ &$v$\\
10013&223&228& $19.33_{-0.13}^{+0.14} $ &$v$\\
34052&225&226& $>20.57$ &$v$\\
51412&225&226& $>20.58$ &     $v$\\
68768&225&226& $>20.53$ &$v$\\
116954&8680&8845& $>21.40$ &$v$\\
184147&3318&3379& $>21.32$ &$v$\\
\gray 4486&49&49& $20.10_{-0.22}^{+0.27} $ &$b$\\
\gray 5918&49&49& $20.20_{-0.30}^{+0.41} $ &   $b$\\
\gray 16711&224&227& $21.06_{-0.24}^{+0.30} $ &$b$\\
\gray 23377&177&179& $21.27_{-0.40}^{+0.65} $ &$b$\\
\gray 40696&197&198& $>21.02$ &$b$\\
\gray 58053&395&396& $>20.98$ &$b$\\
4281&49&49& $>20.39$ &$u$\\
5713&49&49& $>20.22$ &$u$\\
15799&224&227& $>21.89$ &$u$\\
22560&224&227& $>21.57$ &$u$\\
29221&148&149& $>20.19$ &$u$\\
43093&1831&1913& $>22.68$ &$u$\\
60462&1843&1901& $>21.79$ &$u$\\
\gray 4077&49&49& $>19.87$ &$uvw1$\\
\gray 5509&49&49& $>19.27$ &$uvw1$\\
\gray 11768&191&194& $>19.92$ &$uvw1$\\
\gray 21645&220&222& $>21.54$ &$uvw1$\\
\gray 28460&220&222& $>20.90$ &$uvw1$\\
\gray 42229&1852&1937& $>21.28$ &$uvw1$\\
\gray 59599&1864&1925& $>22.37$ &$uvw1$\\
\gray 73712&221&221& $>21.22$ &$uvw1$\\
3872&49&49& $>21.95$ &$uvm2$\\
5304&49&49& $>21.85$ &$uvm2$\\
10916&219&223& $>21.58$ &$uvm2$\\
27553&220&222& $>20.93$ &$uvm2$\\
34905&195&196& $>21.36$ &$uvm2$\\
48416&2051&2142& $>22.44$ &$uvm2$\\
65785&2063&2130& $>21.52$ &$uvm2$\\
\gray 4895&48&49& $>20.19$ &$uvw2$\\
\gray 6304&37&37& $>20.18$ &$uvw2$\\
\gray 33138&221&222& $>21.59$ &$uvw2$\\
\gray 50495&220&221& $>21.24$ &$uvw2$\\
\gray 67853&221&222& $>22.16 $ &$uvw2$\\
\bottomrule
\end{tabular}
\footnotetext{The first column gives the logarithmic
mid-time in seconds after the onset of the GRB. The second and the third 
columns give the start/end of the observations. The data are not
corrected for Galactic extinction.}
\label{UVOT}
\end{minipage}
\end{table}


\begin{table}[thb]
\begin{minipage}[t]{\columnwidth}
\caption{Values plotted in Fig.~\ref{SFD}.}
\centering
\renewcommand{\footnoterule}{}
\begin{tabular}{lcccccrl}
\toprule
Filter & $\lambda$  &  $\nu$(1+z) &   mag  &  $F_\nu$ \\
       & (\AA)      &($10^{15}$Hz)&        &  ($\mu$Jy)\\
\midrule
\gray $R_C$  & 6588 &   2.17  &  $18.83\pm0.04$    & $90.40 \pm3.33$ \\
      $v$    & 5402 &   2.65  &  $20.03\pm0.2$     & $35.38 \pm6.52$ \\
\gray $b$    & 4329 &   3.31  &  $21.28\pm0.3$     & $13.09 \pm3.62$ \\
      $u$    & 3501 &   4.09  &  $>22.13  $        & $ < 2.54 $   \\
\gray $uvw1$ & 2634 &   5.43  &  $>21.33  $        & $ < 2.56 $   \\
      $uvm2$ & 2231 &   6.41  &  $>23.33  $        & $ < 0.41 $   \\
\gray $uvw2$ & 2030 &   7.05  &  $>21.33  $        & $ < 2.82 $   \\
\bottomrule
\end{tabular}
\footnotetext{The $u$, $uvw1$,  $uvm2$ and
$uvw2$ data are upper limits.  For all filters except $R_C$,  $\lambda$ is
from  Poole et al. (\cite{Poole2008}). Data refer to $t$=0.27 days.  A
redshift of $z$=3.773 was assumed. The values have been corrected for
Galactic extinction. The fluxes have been calculated assuming the  conversion
factors from Bessell (\cite{Bessell1979}) and Poole et
al. (\cite{Poole2008}).}
\label{UVOT2}
\end{minipage}
\end{table}


We combined our UVOT data (Table~\ref{UVOT}) with further data reported in the
GRB Coordinates Network Circulars (Rykoff \& Schaefer \cite{Rykoff2006};
Schaefer et al. \cite{Schaefer2006}; Khamitov et al. \cite{Khamitov2006a,
Khamitov2006b}; Malesani et al. \cite{Malesani2006}; Zhai et
al. \cite{Zhai2006}; Karska \& Garnavich \cite{Karska2006} and Sharapov et
al. \cite{Sharapov2006}), all taken in the $R_C$ filter, or unfiltered
calibrated to the $R_C$ band. Karska \& Garnavich (\cite{Karska2006}) point
out zero-point discrepancies between different USNO $R_C$ magnitudes, and we
can confirm that the magnitudes reported by Khamitov et
al. (\cite{Khamitov2006a}) are about one magnitude fainter than what would be
expected from the joint light curve (see below), whereas the late detection by
Khamitov et al. (\cite{Khamitov2006b}) agrees well with the steep decay slope
found by  Karska \& Garnavich (\cite{Karska2006}) and one additional point
from Pozanenko et al., in preparation. We added an error of 0.1
magnitudes in  quadrature to all GCN data points to account for the different
filters and reference stars.

\begin{figure}
\centering
\includegraphics[width=8.0cm]{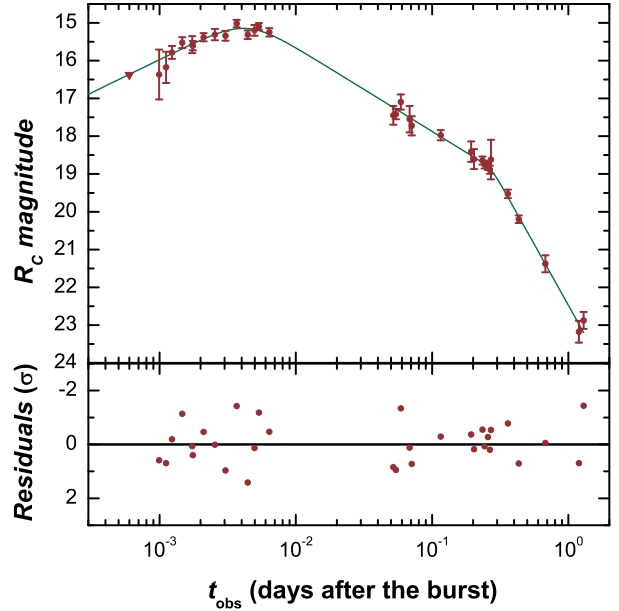}
\caption{
The composite $R_C$-band light curve of the afterglow of GRB 060605 (see text 
for more details), fitted with a double broken power-law. 
The lower panel gives the residuals of the fit. These data are
corrected for Galactic extinction. See the text for the results of the fit.}
\label{opticalLC}
\end{figure}

Using the $R_C$-band light curve as the most reliable template, and correcting
all data for the foreground extinction of $E_{B-V}=0.049$ (Schlegel et
al. \cite{Schlegel1998}), we derive colours for the UVOT detections. We find
$v-R_C=1.2\pm0.2$, $white-R_C=2.3\pm0.2$, $b-R_C=2.45\pm0.3$, $u-R_C>3.3$,
$uvw1-R_C>2.5$, $uvm2-R_C>4.5$, and $uvw2-R_C>2.5$. We note that, usually,
$white$ magnitudes are close in value to UVOT $v$-band values. Given the high
redshift of the source, however, the large $white-v$ colour is due to the
unfiltered UVOT bandpass being strongly affected by Lyman damping, making the
afterglow much redder than usual (see below). We used the derived colour
indices to shift the UVOT detections ($v$, $white$ and $b$) to the $R_C$-band
and to construct a composite light curve (Fig.~\ref{opticalLC}).

In the UV/optical bands the data are broadly consistent with an achromatic
evolution,  but we caution that the data are sparse.  We find an early rise,
as reported by  others (Schaefer et al. \cite{Schaefer2006}; Zhai et
al. \cite{Zhai2006}), which is followed by a ``classical'' broken power-law
decay. Denoting the three slopes $\alpha_R$ (where the index $R$ stands for
\it rise\rm), $\alpha_1$ and $\alpha_2$, we find $\alpha_R=-0.70\pm0.15$,
$\alpha_1=0.89\pm0.04$ and $\alpha_2=2.58\pm0.15$. The break times are
$0.0044 \pm 0.0007$ days for the break from rise to decay, and $0.27\pm0.02$
days for the second break. In both cases, we assumed that the  host galaxy
underlying the afterglow has an extinction-corrected magnitude of  $R_C=26.3$
(see Sect.~\ref{host}). We fixed the break smoothness parameter $n_2$
according to Liang et al. (\cite{Liang2008}) to 10 for the second break. For
the first break, while we were not able to leave $n_1$ as a free parameter of
the fit, we find a minimum $\chi^2$ and a very good fit ($\chi^2/{\rm
d.o.f.}=21.00/32=0.66$) for a rather smooth break $n_1=2.5$.  A summary of the
fit parameters is given in Tab.~\ref{tab:optical}

\begin{table}
\begin{minipage}[t]{\columnwidth}
\caption{Parameters of the fit of the optical afterglow light curve
(Fig.~\ref{opticalLC}).}
\centering
\renewcommand{\footnoterule}{}
\begin{tabular}{ll}
\toprule
Parameter	& Value\\
\midrule
\gray $\alpha_{\rm R}$	& $-0.70 \pm 0.15$\\
$\alpha_1$		& $ 0.89 \pm 0.04$\\
\gray $\alpha_2$	& $ 2.58 \pm 0.15$\\
$t_{\rm 1} \ \rm (days)$& $0.0044 \pm 0.0007$\\
\gray $t_2 \ \rm (days)$& $0.27 \pm 0.02$\\
$n_1$			& 2.5\\
\gray $n_2$		& 10\\
$\chi^2/{\rm d.o.f}$	& $21.00/32=0.66$\\
\bottomrule
\end{tabular}
\footnotetext{The slope $\alpha_R$ describes the rising part of the light curve,
$(t_1, n_1$) and $(t_2, n_2)$ describe the two breaks.}
\label{tab:optical}
\end{minipage}
\end{table}

The peak time of $358\pm61$ s ($0.0041 \pm 0.0007$ days) can be found from the
light curve fit  by setting  d$F_\nu(t)$/d$t = 0$ and has a value that is
comparable to what has been found  for, e.g., the early phase of the optical
afterglow of GRB 060418 and 060607A (Molinari et al. \cite{Molinari2007}). Our
result for $\alpha_1$ is in agreement with the value reported by Schaefer et
al. (\cite{Schaefer2006}) and the peak time we derive is in agreement with
Zhai et  al. (\cite{Zhai2006}).

The afterglow of GRB 060605 belongs to the growing ensemble of optical
afterglows for which thanks to a rapid response  in the follow-up observations
the data show the early rise  of the afterglow, as predicted by theoretical
models (Panaitescu \& Kumar \cite{Panaitescu2000}; Sari \cite{Sari1997}). With
a peak magnitude of $R_C$ = 15.2 at $t\approx 360$ s (0.0042 days)
(Fig.~\ref{opticalLC}) it is among the brightest optical afterglows ever
detected (Nardini et al. \cite{Nardini2008}; Kann et al. \cite{Kann2007}).

\begin{figure}
\centering
\includegraphics[width=8.0cm]{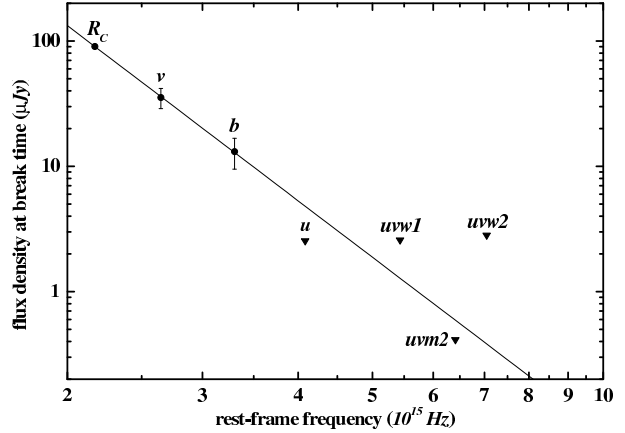}
\caption{Spectral energy distribution (SED) of the afterglow of GRB
060605 in the optical bands
at $t$=0.27 days. The line is the best-fit power law.
Detections in three filters ($bvR_C$) and deep upper limits in $u$ and 
in $uvm2$ show a very steep SED ($\beta_{\rm opt}= 4.64\pm0.58$), a result of 
Lyman blanketing in the optical/UV due to the high redshift.}
\label{SFD}
\end{figure}

Using the colours derived above, we can construct the spectral energy
distribution (SED; Fig.~\ref{SFD} and Table~\ref{UVOT2}) at $t$=0.27 days.  We
have detections in only three filters ($BVR_C$). These three data points can
be fit with a very steep SED: we find $\beta_{\rm opt}= 4.64\pm0.58$. The
steep  slope is further confirmed by the $u$- band and $uvw2$-band upper
limits, the other two filters are less  constraining. This is much steeper
than typical afterglow slopes, which lie in the range from 0.5 to 1.1 (e.g.,
Kann et al. \cite{Kann2006,Kann2007}).  This is mostly due to the Lyman
blanketing blueward of the rest-frame  Ly $\alpha$ line, which at $z = 3.773$
falls between the observed $v$ and $R_C$  bands.

\section{Discussion}

\subsection{The X-ray light curve \label{Xcurve} }

According to Nousek et al. (\cite{Nousek2006}) and Zhang et
al. (\cite{Zhang2006}), the X-ray light curve of an afterglow has a canonical
shape. It consists of four well-defined evolutionary phases: a steep initial
decay followed by a flat (plateau) phase,  then a steeper decay  (a pre-jet
break phase)  and finally a post-jet break phase (e.g., GRB 050315; Vaughan et
al. \cite{Vaughan2006}; Panaitescu \cite{Panaitescu2008}; Liang et
al. \cite{Liang2008}).  However, as we have already emphasized, as a first
guess the data  seem to imply that the X-ray light curve of GRB 060605
consists of three power-law  segments only (Fig.~\ref{fig:lc}). Let us first
discuss if this allows a satisfactory explanation of the observational data
within the context of the fireball model. For this reason, let as  discuss the
nature of the second break seen in the fitted X-ray light curve at around
0.09 days. Figure~\ref{fig:lc} suggests to consider two cases.

\emph{Case A:} If this is the jet break, segment II of our  Fig.~\ref{fig:lc}
is the pre-break segment, while segment III is the post-break phase.
Consequently  the slopes of segment II ($\alpha_{\rm II}$) and of segment III
($\alpha_{\rm III}$)  are the pre-jet break decay  and the post-break decay
slopes, usually designated as $\alpha_1$ and $\alpha_2$, respectively.

\emph{Case B:} The second possibility is that the jet break was in fact at much
later times, i.e. at $\gtrsim  1.2$ days and $\alpha_1=\alpha_{\rm III}$.
Indeed, most X-ray afterglows do not show jet breaks (Liang et
al. ~\cite{Liang2007}), with the most extreme example being GRB 060729 (Grupe
et al. ~\cite{Grupe2007b}).

If indeed one of these two possibilities is compatible with the  theoretical
framework can be discussed first based  on the $\alpha-\beta$ relations
(cf. Zhang \& M\'esz\'aros 2004).  For this reason, we  considered  the
standard wind and ISM models for the isotropic case as well as for a jet with
the cooling frequency $\nu_c$ below and above the observers window. In
addition, we also considered models with a power-law index of the electron
distribution function of less than 2, as they are  listed in table 1 of Zhang
\& M\'esz\'aros (\cite{Zhang2004}). Based on our data the latter models
turned out to be excluded with high significance, however. So, we followed
Greiner et al. (\cite{Greiner2003}, their table 6) and present here the
results for the eight standard cases.

Table~\ref{alphabeta} shows the predicted values for  the spectral slope
$\beta$ for the considered scenarios as a function of the observed light curve
decay slope in the X-ray band. These results have to be compared with the
observed $\beta$ in the X-ray band, $\beta_{\rm X} = 1.06\pm0.16$ (Sect.
\ref{xray}). For Case B we calculated $\alpha_2$ via $\alpha_2 = \alpha_1 +
1.14$, with the latter being the mean value for $\Delta \alpha$ in our  data
base of GRB afterglows  with an error of 0.3 to be very
conservative. Table~\ref{alphabeta} shows that Case A is ruled out with high
significance: it predicts a spectral slope before the break time (i.e., for
the isotropic case) which is in clear disagreement with the observational
data. This holds for the slow as well as for the fast cooling case.  Case B
suggests that in the X-ray band the afterglow was in the slow-cooling regime,
and the wind as well as the ISM model is in agreement with the data.  
However, as we will discuss
in the following, even this case is disfavoured if one considers other
theoretical criteria.

\begin{table}[t]
\begin{minipage}[t]{\columnwidth}
\caption{Predicted spectral slope $\beta$ in the X-ray band
for the two different cases A and B discussed in Sect.~\ref{Xcurve}.} 
\centering
\renewcommand{\footnoterule}{}
\begin{tabular}{clcc}
\toprule
{\bf Model}  & {\bf $\beta(\alpha)$}   &{\bf $\beta_A$}& {\bf $\beta_B$} \\
\midrule
Fast cooling ($\nu_c < \nu_{\rm X}$) &&& \\
\midrule
\gray ISM$_{iso}$ & $(2\alpha_1+1)/3$& 0.56$\pm$0.02 & 1.59$\pm$0.05\\ 
\gray ISM$_{jet}$ & $\alpha_2/2$     & 0.94$\pm$0.04 & 1.52$\pm$0.15\\
wind$_{iso}$ & $(2\alpha_1+1)/3$     & 0.56$\pm$0.02 & 1.59$\pm$0.05\\
wind$_{jet}$ & $\alpha_2/2$          & 0.94$\pm$0.04 & 1.52$\pm$0.15\\
\midrule
Slow cooling ($\nu_c > \nu_{\rm X}$) &&& \\
\midrule  
\gray ISM$_{iso}$ & $ 2\alpha_1/3$   & 0.23$\pm$0.02  & 1.26$\pm$0.05\\
\gray ISM$_{jet}$ & $(\alpha_2-1)/2$ & 0.44$\pm$0.04  & 1.02$\pm$0.15\\
wind$_{iso}$  &(2$\alpha_1-1)/3$     &$-$0.11$\pm$0.02& 0.93$\pm$0.05\\
wind$_{jet}$  & $(\alpha_2-1)/2$     & 0.44$\pm$0.04  & 1.02$\pm$0.15\\
\bottomrule
\end{tabular}
\footnotetext{The calculated spectral slopes $\beta_A$, and $\beta_B$ have to be 
compared with the observed $\beta_{\rm X} = 1.06\pm0.16$ (Sect.~\ref{xray}).
For Case A it is $\alpha_1 = 0.34\pm$0.03 and $\alpha_2= 1.89\pm$0.07,
while for Case B it is $\alpha_1 = 1.89\pm$0.07 and $\alpha_2 = 3.03\pm$0.30.}
\label{alphabeta}
\end{minipage}
\end{table}

\subsection{The broad-band SED \label{broadSED} }

\begin{figure}
\includegraphics[angle=0,width=8.5cm]{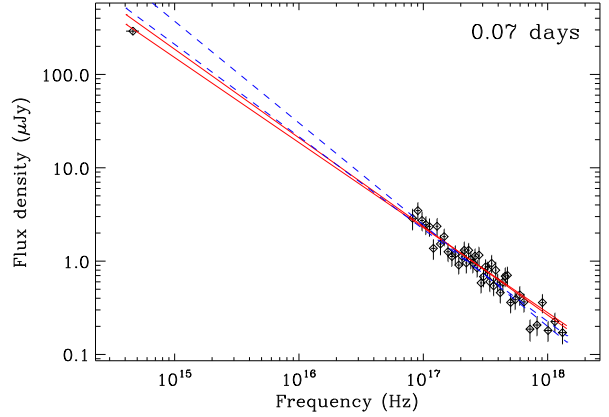}
\caption{
Broad-band spectrum of the afterglow of GRB 060605 at 0.07 days. The lines
show the 1-$\sigma$ confidence regions of power-law fits. Solid lines:
simultaneous fit of  the X-ray and the $R_C$-band data; dashed lines: fit of
the X-ray data only. The $R_C$-band data point was calculated from the UVOT
data (Sect.~\ref{optcurve}).  It lies beneath the spectral slope extrapolated
from the X-ray band, indicating  that the cooling frequency is in between the
optical and the X-ray band.}
\label{OXSED007}
\end{figure}

\begin{figure}
\includegraphics[angle=0,width=8.5cm]{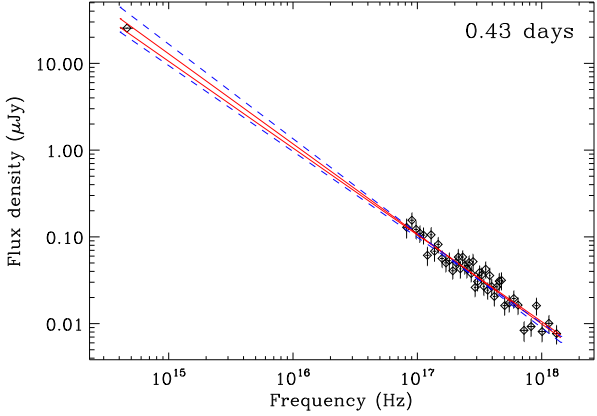}
\caption{
The same as Fig.~\ref{OXSED007} but at 0.43 days. Here,  the  $R_C$-band data
point is from Sharapov et al. (\cite{Sharapov2006}). A  simultaneous fit of
the X-ray and  the $R_C$-band data gives the same result as a fit of the X-ray
data only, suggesting a change of the SED compared to 0.07 days
(Fig.~\ref{OXSED007}).}
\label{OXSED}
\end{figure}

In order to better distinguish among possible scenarios for the afterglow
emission of GRB 060605 in the context of the standard fireball model, we
studied the broad-band spectrum from the X-rays (0.3--6 keV) to the optical
($R_C$ band) at two epochs. The first epoch, at 0.07 days
(Fig.~\ref{OXSED007}), was chosen because it is before any suspected X-ray jet
break time, while the second epoch, at 0.43 days (Fig.~\ref{OXSED}),
corresponds to the time of the $R_C$-band measurement by  the Nordic Optical
Telescope (NOT; Sharapov et al.~\cite{Sharapov2006}).

The $R_C$-band magnitude at 0.07 days was derived from UVOT data with the
procedure described in Sect.~\ref{optcurve}. In order to check the
reliability of this method, we computed in the same way the $R_C$ magnitude at
0.43 days, finding a value fully consistent with the NOT measurement. The
$R_C$-band fluxes were corrected for extinction in our Galaxy. The 0.3 to
6 keV spectra were derived in the following way: we deconvolved the average
XRT count spectrum by assuming the best-fit power-law model, corrected it for
the  measured column density (Sect.~\ref{xray}), and rescaled it to the two
epochs by  using the multi-broken power-law model that fitted the XRT light
curve best. The fit of the X-ray spectrum alone, with $N_{\rm H}$ fixed at the
Galactic value of $5.1\times10^{20}$ cm$^{-2}$, gives a spectral index of
1.04$\pm$0.05. Both, X-ray and $R_C$-band fluxes, were converted into flux
densities ($\mu$Jy), in order to build-up the SED and to allow for broad-band
spectral fitting.

As it can be seen in Figs.~\ref{OXSED007} and \ref{OXSED}, there is a hint of
a spectral evolution between the two epochs. At 0.43 days the $R_C$-band flux
density is fully consistent with the extrapolation of the power-law fit of the
X-ray  spectrum. This is confirmed by a fit of the SED from the optical
($R_C$-band) to the X-ray band with a simple power-law,  which provides an
acceptable chi-square value (39.5/44=0.90) and a spectral index of $\beta_{\rm
opt,X}= 1.02\pm$0.02. This is evidence that at this epoch the cooling
frequency, $\nu_c$, was already lower than the $R_C$-band one. On the other
hand, in the spectrum at 0.07 days the $R_C$-band data point is below the
extrapolation of the power-law fit of the X-ray spectrum. Thus, at 0.07 days
the cooling frequency was still at slightly higher frequencies than the $R_C$
band, suggesting that in  both spectra $\nu_c$ is below the X-ray band  (i.e.,
fast cooling regime at X-rays) and decreasing with time. The latter points to
an ISM environment (Sari et al. 1998), since for a wind model  $\nu_c$ is
increasing with time (Chevalier \& Li 2000).  The former excludes even Case B
(Sect. \ref{Xcurve}), since the $\alpha-\beta$ relations are not in agreement
with the fast cooling regime in the X-ray band.  Moreover, given that at
$t$=0.43 days the $R_C$-band data point of the afterglow light curve  lies
exactly on the SED derived in the X-ray band (with a slope of $\beta_{\rm X} =
1.02 \pm 0.02$; see Fig.~\ref{OXSED}), the power-law index $p$ of the electron
distribution function is $p=2\beta=2.04 \pm 0.04$, a value close to the
observed mean (cf. Kann et al. \cite{Kann2006};  see also Starling et
al. \cite{Starling2008}).  If the cooling frequency were  at higher values
than  the X-ray band ones, then $p=2\beta + 1=3.04 \pm 0.04$, an unusually
large number. Consequently, the data  disfavour Case B and thus  the
hypothesis of a jet break occurring after about 1.2 days.

To summarize, neither Case A or Case B lead to a reasonable agreement with the
theoretical framework. While one could argue that this points to  a problem
with the theory, we suggest that the most reasonable hypothesis is to assume
that in fact our ansatz of a three-segment X-ray light curve is incorrect and
a forth power-law segment is needed.  This hypothesis is motivated by the
observations that  many, if not most, X-ray light curves can be described  by
a canonical shape (Nousek et al. \cite{Nousek2006}; Zhang et
al. \cite{Zhang2006}). In fact, the existence of a 4th power-law segment in
the X-ray light curve is also supported by an F-test. It shows that a fit of
the X-ray light curve improves if actually  two breaks exist after 0.004 days
instead of just one, and that the 4th power-law segment is placed between
about 0.07 and 0.27 days. The  goodness of fit, $\chi^2/$d.o.f., improves from
54.21/39=1.39 to 50.74/37=1.18 if an  analytical ansatz is made that allows
for the occurrence of two breaks after 0.004 days. This translates into a
significance for an additional break of $2.3\,\sigma$, or a 2\% probability to
find such an additional break by chance. In order to improve the fit
further, we finally performed a joint fit of the optical and X-ray data.


\subsection{The X-ray vs. the optical light curve: a joint fit \label{XO}}

\begin{figure*}
\centering
\includegraphics[width=15.0cm,clip=true]{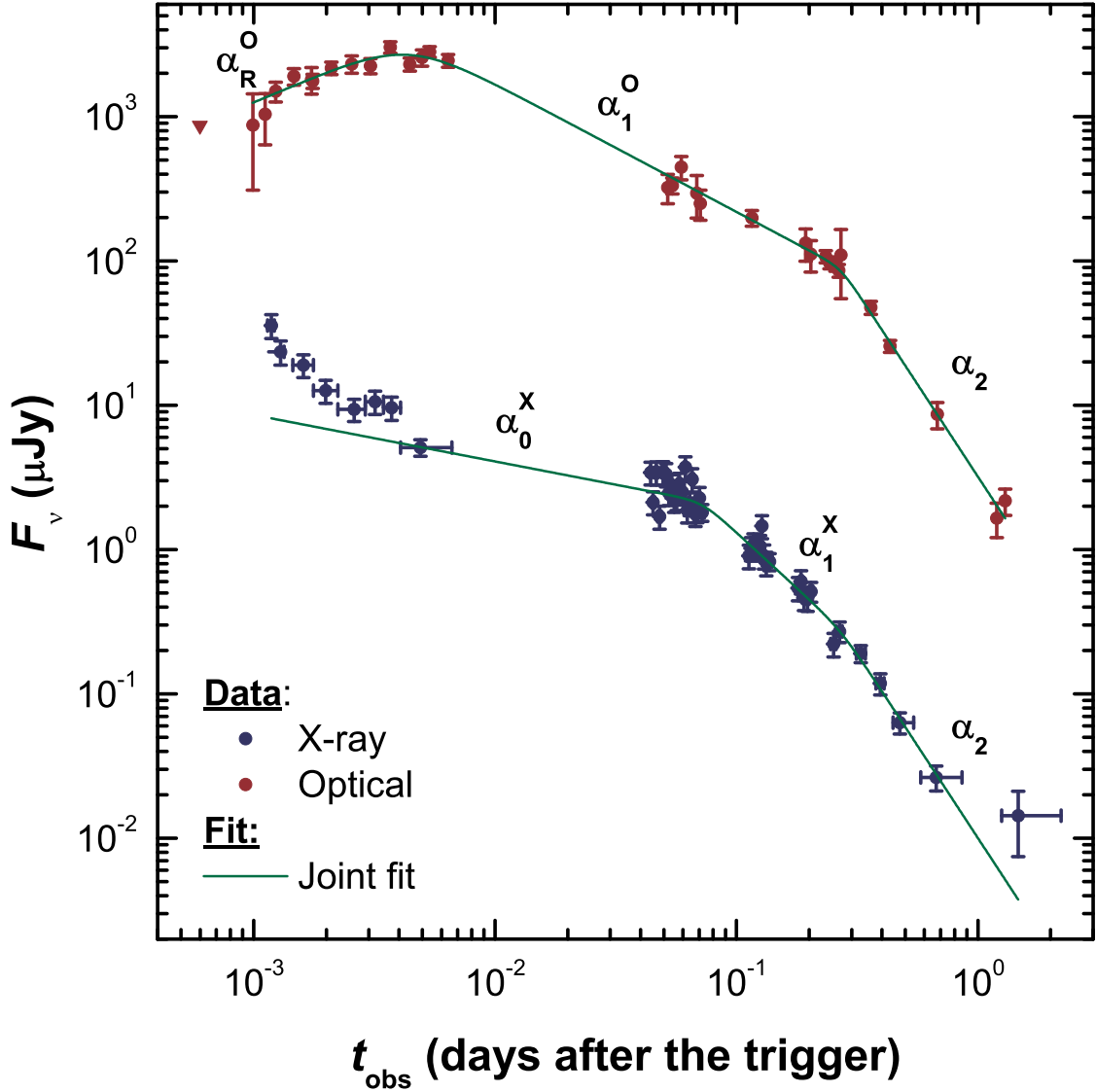}
\caption{
The composite optical and X-ray light curve of the afterglow of GRB 060605.
The final X-ray point is an upper limit.  The green line shows the
results of the joint fit of the optical  and the X-ray light curve (see
Sect.~\ref{XO}). Note that the joint fit included only the X-ray data from
$t>$0.004 days.}
\label{fig:joint}
\end{figure*}

Figure~\ref{fig:joint} shows the combined optical/X-ray light curve of the
afterglow of GRB 060605. At early times, from about 0.0012 to 0.0046 days, the
X-ray and the optical light curves show a completely different behaviour. The
X-ray light curve is falling while the optical light curve is rising, similar
to what was observed for  e.g. GRB 060418 (Jin \& Fan \cite{Jin2007}). This
rising optical component ends approximately at  the same time as the plateau
phase commences in  X-ray band. The optical light curve might also include a
plateau phase lasting for at least 100 s (0.0012 days) around the peak time.

The later behaviour of the light curve is difficult to interpret with
certainty due to the lack of X-ray as well as optical data between about 0.006
and 0.041 days. Potentially, also the optical light curves could show a
long-lasting plateau phase in this period if its peak was followed by a faster
decay. But in this case there are no data published to check this hypothesis.

In doing the joint fit, we set the following boundary conditions: (a) an
identical decay slope of the optical and the X-ray light curve and (b) an
identical jet break time. In addition, as discussed before, the numerical
ansatz included two breaks in the X-ray light curve after 0.004 days. For the
fit the optical data of Khamitov et al. (\cite{Khamitov2006a}) have been
excluded  since they are roughly one magnitude too faint (Karska \& Garnavich
\cite{Karska2006}). The data from 0.05 to 0.1 days are the UVOT $white$, $b$
and $v$ measures shifted to the $R_C$-band zero point using the early $R_C$
and UVOT observations, and thus may be incorrect if a strong colour change
occurred in between. However, no sign of a strong chromatic evolution is
detected.

Allowing for a different pre-break decay slope in the optical  and in the
X-ray band, the joint fit finds a break time of $0.27 \pm 0.02$ days, a
pre-break decay slope in the optical of $0.89 \pm 0.04$, a pre-break decay
slope in the X-ray band of $1.54 \pm 0.11$, and a post-break decay slope of
$2.56\pm 0.13$ (Fig.~\ref{fig:joint} and Tab.~\ref{tab:joint1}).  During the
fit, the smoothness parameters were always fixed. Applying the $\alpha-\beta$
relations now shows that a wind model is disfavoured. This is in agreement
with our results in Sect.~\ref{broadSED}, where we found that the cooling
frequency was decreasing with time, indicating in this way  an expansion of
the fireball into a circumburst medium with an ISM profile. A decision between
fast and slow cooling cannot be made, however. Both are acceptable within
2$\sigma$.  Given that we know already from our analysis of the SED that  the
X-ray emitting electrons were in the fast cooling region at 0.07 days after
the burst (Sect.~\ref{broadSED}), one might wonder if an agreement between
theory and observations at just the $2\sigma$ level is acceptable at all.  We
believe that the data quality is simply not good enough in order to obtain
more robust results. At least it is obvious that the joint fit in combination
with a 4-segment X-ray light curve provides a much better explanation of the
observational data than a simple 3-segment ansatz (Sect.~\ref{Xcurve}).

\begin{table}
\caption{Fit parameters of the joint optical and X-ray fit 
($\chi^2/\rm{d.o.f.}=62.11/65=0.96$).}
\centering
\begin{tabular}{ll}
\toprule
\multicolumn{2}{c}{\textbf{Before the jet break}}\\
Parameter                                 & Value\\
\midrule
\gray $\alpha^{\rm opt} _R$               & $-0.70 \pm 0.15$\\
      $\alpha^{\rm opt} _1$               & $0.89  \pm 0.04$\\
\gray $t^{opt} _{\rm{b},\,1}$ (days)      & $0.004\pm 0.0007$\\
      $n^{\rm opt} _1$                    & 2.5\\
\gray $\alpha^{\rm X}_0$                  & $0.32 \pm 0.06$\\
      $\alpha^{\rm X}_1$                  & $1.54  \pm 0.11$\\
\gray $t^{X} _{\rm{b},\,1}$ (days)        & $0.072 \pm 0.008$\\
      $n^{\rm X} _1$                      & 10\\
\midrule
\multicolumn{2}{c}{\textbf{After the jet break}}\\
Parameter                                 & Value\\
\midrule
\gray $\alpha _2$                         & $2.56 \pm 0.13$\\
      $t_{\rm{b},2}\,\rm \left( days\right)$& $0.27 \pm 0.02$\\
      $n_2$                               & 10\\
\bottomrule
\end{tabular}
\label{tab:joint1}
\end{table}

\begin{table}
\caption{
Predicted spectral slope $\beta$  considering the parameters  listed 
in Tab~\ref{tab:joint1}
for different applied models. The spectral slopes have to be  compared with
the observed $\beta_{\rm X} = 1.06\pm0.16$ (Sect.~\ref{xray}).}
\centering
\begin{tabular}{lll}
\toprule
\textbf{Model}      & $\beta\left(\alpha\right)$   &  value \\
\midrule
\multicolumn{3}{l}{Fast cooling}\\
\midrule
\gray ISM$_{iso}$   & $(2\alpha^{\rm X} _1 +1)/3$  & $1.36\pm0.07$\\
\gray ISM$_{jet}$   & $\alpha_2 /2$                & $1.28\pm0.07$\\
wind$_{iso}$        & $(2\alpha^{\rm X} _1 +1)/3$  & $1.36\pm0.07$\\
wind$_{jet}$        & $\alpha_2 /2$                & $1.28\pm0.07$\\
\midrule
\multicolumn{3}{l}{Slow cooling}\\
\midrule
\gray ISM$_{iso}$   & $2\alpha^{\rm X} _1 / 3$     & $1.03\pm0.07$\\
\gray ISM$_{jet}$   & $(\alpha_2 -1)/2$            & $0.78\pm0.07$\\
      wind$_{iso}$  & $(2\alpha^{\rm X}_1 -1)/ 3$  & $0.69\pm0.07$\\
      wind$_{jet}$  & $(\alpha _2 -1)/ 3$          & $0.78\pm0.07$\\
\bottomrule
\end{tabular}
\label{tab:joint2}
\end{table}

\begin{figure}
\centering
\includegraphics[width=8.0cm,clip=true]{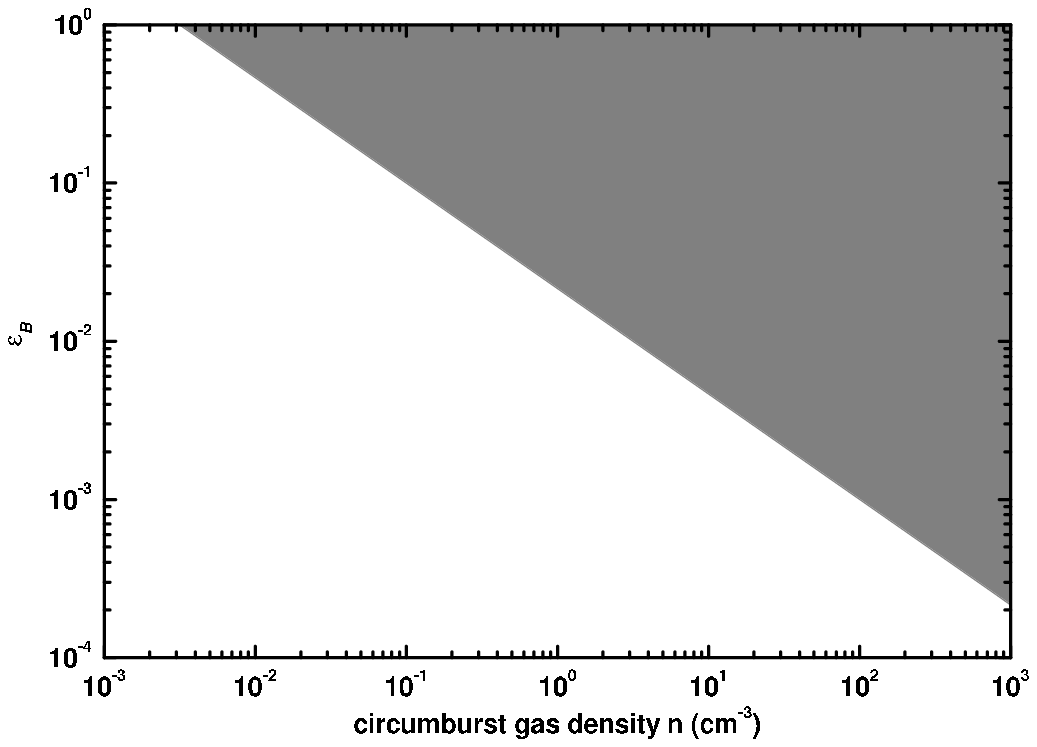}
\caption{
Constraints on the parameter space of  the fractional energy carried by the
magnetic field, $\varepsilon_{B}$, and the gas density (in units of cm$^{-3}$)
in the circumburst medium into which the fireball was expanding. It was
assumed here that at 0.43 days the cooling frequency was at frequencies less
than $4.5\,\times 10^{14}$ Hz ($R_C$ band). The allowed region is represented
by the grey area.}
\label{ism}
\end{figure}

We can now use the observational data to constrain the density  $n$ in
the circumburst medium and the parameter $\varepsilon_B$ that measures the
fraction of energy carried by the magnetic field. For an ISM medium the cooling
frequency is given by (cf. Granot et al. 2000)
\begin{equation}
\nu_c = 2.9\,\times\,10^{15}\ 
    \varepsilon_{B, -1}^{-3/2} E_{52}^{-1/2} n_0^{-1} t_2^{-1/2}
    (1+z)^{-1/2}\ \ \mbox{Hz}\,,
\label{epsilo}
\end{equation}
where $n_0 = n/$1 cm$^{-3}$, $\varepsilon_{B, -1}= \varepsilon_{B}/0.1$, and
$t_2=t/100$ s.  Using $\nu_c < 4.5\,\times\,10^{14}$ Hz at $t$=0.43 days with
$E_{52}=2.5$ (Butler et al. \cite{Butler2007}) and  $z$=3.773  we can
constrain the product $\varepsilon_{B, -1}^{-3/2} n_0^{-1}$
(Fig.~\ref{ism}). Since $\varepsilon_{B}<1$, it must be $n \gtrsim 0.005$
cm$^{-3}$, a reasonable result. On the other hand, the low deduced hydrogen
column density along the line of sight in the GRB host
(Sects.~\ref{NHoptical}, \ref{xray}) might indicate a  relatively low
circumburst gas density.  If we require $n < 100$ cm$^{-3}$ then
$\varepsilon_{B} > 0.001$, which is also a reasonable constraint.

In the following we consider the break at 0.27 days as a classical jet break
and we use the results of the  joint fit to discuss the energetics of the
afterglow.

\subsection{Energetics \label{jets}}

\subsubsection{The burst \label{Egamma}} 

We follow the standard approach to calculate the jet  half-opening angle for
an ISM environment (cf. Sari at al. \cite{Sari1999}),
\begin{equation}
\Theta^{\rm ISM}_{\rm jet} = 
\frac{1}{6}\left(\frac{t_b}{1+z}\right)^{3/8}\left(\frac{n_0 \eta_\gamma}
{E_{52}}\right)^{1/8} \ \mbox{rad}\,.
\label{sari}
\end{equation}
Here, $E_{52}$ is the isotropic equivalent energy of the prompt emission in
units of $10^{52}$ erg, $n_0$ is the density of the ambient medium in
cm$^{-3}$, $\eta_\gamma$ is the efficiency of the shock in converting the
energy of the ejecta into gamma radiation, and $t_b$ is the break time in
days. We set $n_0=1$ cm$^{-3}$ and $\eta_\gamma=0.2$. Assuming the observed
break time at $t_b$= 0.27$\pm$0.02 days, as follows from the joint fit, with
$E_{52}=2.5^{+3.1}_{-0.6}$ (Butler et al. \cite{Butler2007}) we get
$\Theta^{\rm ISM}_{\rm jet} = 2.37_{-0.10}^{+0.37}$ degrees. This can be
compared with a mean value of the half-opening angle of the pre-\emph{Swift}
era GRBs of 4 degrees (Zeh et al. \cite{Zeh2006}) with a width of 0.13 dex.
So, the jet of GRB 060605 was narrowly beamed but the value found is not
extraordinary.  Changing $\eta_\gamma$ to 1.0 does not increase $\Theta^{\rm
ISM}_{\rm jet}$ in a notable manner due to the weak dependence of it on
$\eta_\gamma$. On the other hand, as already  stated before (see
Eq.~\ref{epsilo}), a very high gas density seems to be unlikely given that  we
do not see so much hydrogen  at the redshift of the burst in the X-ray
spectrum and in the optical spectrum as well. Assuming the above numbers, the
corresponding beaming-corrected energy release in the gamma-ray band is
$E_\gamma^{\rm corr} = 2.14^{+2.76}_{-0.52}\,\times\,10^{49}$ erg. This is
among the  smallest $E_\gamma^{\rm corr}$ found so far but, again,  not
exceptionally small (cf. Zeh et al. 2006; Racusin et
al.~\cite{Racusin2008}). The small $\Theta^{\rm ISM}_{\rm jet}$ and
$E_\gamma^{\rm corr}$ match into the picture (Racusin et
al.~\cite{Racusin2008}), according to which \emph{Swift} GRBs have on average
smaller jet opening angles and lower collimated $\gamma$-ray energies than
pre-\emph{Swift} bursts (which were on average at lower redshift).

As outlined by Panaitescu \& Kumar (\cite{Panaitescu2000}),  assuming that the
observed peak in the optical light curve  signals the fireball deceleration
timescale (which is $t_{\rm peak}/(1+z)$), one can calculate the initial
Lorentz factor, $\Gamma_0$, of the outflow. Following Sari et
al. (\cite{Sari1999}; Eq.~\ref{sari}), in the afterglow deceleration phase the
time evolution of the Lorentz factor is given by
\begin{equation}
\Gamma(t) = 6\,\Big(\frac{E_{\rm 52}}{\eta_\gamma\, n_0}\Big)^{1/8}\, 
\Big(\frac{t}{1+z}\Big)^{-3/8}\,.
\end{equation}
Setting $E_{52}=2.5$ (Butler et al. \cite{Butler2007}), $\eta_\gamma=0.2$, and
$n_0$=1 cm$^{-3}$, it follows $\Gamma (t=t_{\rm peak}) = 116 \pm 15$, and hence
$\Gamma_0 =232 \pm 15$. Following the procedure outlined in  Molinari et
al. (\cite{Molinari2007}) leads to $\Gamma_0 = 366 \pm 23$. These values
are comparable to those of other bursts (cf. Molinari et al.).

\subsubsection{The X-ray afterglow}

The luminosity of the afterglow is (e.g., Nousek et al.~\cite{Nousek2006})
\begin{equation}
L_X(t_{\rm host}) = 4\pi d_L^2\, (1+z)^{\beta-1}\, F_X(t_{\rm obs})\,,
\end{equation}
where $F_X(t_{\rm obs})$  is the observed time-dependent flux in the X-ray
band and $d_L$ is the luminosity distance. Using $z$=3.773 and assuming
$\beta=1.06$ (Sect. \ref{xray}) even at very early times, we get for the time
evolution of the X-ray luminosity of the afterglow  in the 0.3-10 keV energy
band (in units of erg s$^{-1}$)
\begin{eqnarray}
L_{\rm X}(t) &=& 7.0\,\times\,10^{48}\, (t/t_1)^{-\alpha}\,,\
\alpha= 0.32; \ \ t_1 \le t \le t_2,\\
L_{\rm X}(t) &=& 2.0\,\times\,10^{48}\, (t/t_2)^{-\alpha}\,,\
\alpha= 1.54; \ \ t_2 \le t \le t_3,\\
L_{\rm X}(t) &=& 2.6\,\times\,10^{47}\, (t/t_3)^{-\alpha}\,,\
\alpha= 2.56; \ \ t \ge t_3\,,
\end{eqnarray}
where $t=t_{\rm host}$ is measured in the GRB host frame and all the break
times $t_1$, $t_2$ and  $t_3$ are also given in the host frame.  Note that for
reasons of simplicity we have replaced here the complicated light curve fit
of the afterglow with a multiple broken power-law  by three single power-law
decays in between the deduced break times, with $t_1 = 210 s/(1+z) = 44$ s
(Sect.~\ref{xray}), $t_2 =0.072$ days/$(1+z)=0.015$ days
(Tab.~\ref{tab:joint1}), and  $t_3 = 0.27$ days/$(1+z)=0.056$ days
(Tab.~\ref{tab:joint1}).  Since the breaks in the light curve are quite sharp,
this  is surely justified. The behaviour at $t<$44 s (host frame)  we have
excluded here since it might not represent afterglow light.  Based on these
numbers we find that  the isotropic energy release of the afterglow in the
X-ray band was $4.1 \times 10^{52}$ erg between 0.0005 and 0.015 days (which
is 16\% of $E_{\gamma}$; Butler et al. \cite{Butler2007}), $2.5 \times
10^{51}$ erg between 0.015 and 0.056 days (0.10 $E_{\gamma}$), and $8.1 \times
10^{50}$ erg thereafter (0.03 $E_{\gamma}$), assuming a constant decay.

\subsubsection{The optical afterglow}

\begin{figure}
\centering
\includegraphics[width=9.0cm,clip=true]{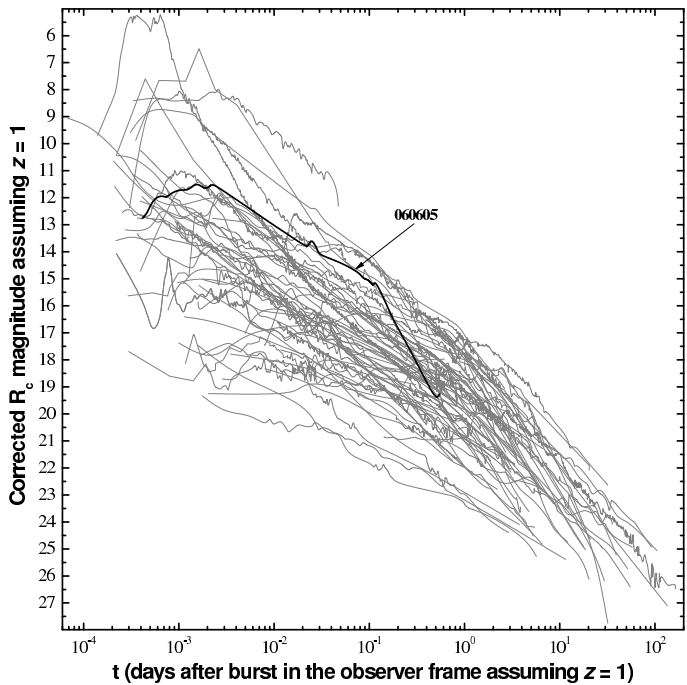}
\caption{
The observed $R_C$-band light curve of the afterglow of GRB 060605 compared to
the ensemble of optical afterglows known so far after shifting all light
curves to a common redshift of $z$=1.  All data are corrected for Galactic
extinction.  See Kann et al. (\cite{Kann2006}) for  the method and Kann et
al. (\cite{Kann2007}) for more details on other bursts.}
\label{z1}
\end{figure}

Similar to the ``Bronze Sample'' of Kann et al. (\cite{Kann2007}), we can
assume that the $R_C$-band afterglow of GRB 060605  is not affected by host
galaxy extinction (which seems to be low at high redshifts anyway, Kann et
al. \cite{Kann2007}). This assumption is also supported by the  observed SED
of the afterglow at 0.43 days (Fig.~\ref{OXSED}).  Furthermore, if the cooling
break $\nu_c$ lies  at wavelengths longer than the optical ones then
$\beta_{\rm opt}=\beta_{\rm X}$. Therefore, assuming $A_V$(host)=0,
$\beta_{\rm opt}=1.06$, and  using the method presented in Kann et al.
(\cite{Kann2006}), we are able to derive  a lower limit on the magnitude shift
$dR_c \ge 3.61$ mag. This shift (see Kann et al. \cite{Kann2006} for more
details) describes the magnitude change that appears  when the afterglow light
curve is corrected for extinction (which we are  unable to do here, therefore
we derive only a lower limit) and shifted to $z=1$ (which also implies a
temporal shift). Comparing the afterglow with the sample presented in Kann et
al. (\cite{Kann2007}), we find that it is among the brightest afterglows at
early times, comparable to the afterglow of GRB 050820A (Fig.~\ref{z1}). At 43
s in the rest-frame ($z$=1 assumed), it has $R_C \le 11.83\pm0.15$, which
places it among the tight clustering found by Kann et al. (\cite{Kann2007}),
although the afterglow is still rising. To derive a magnitude at one day after
the GRB (if at $z$=1), we need to extrapolate the late steep decay. We find
$R_C \le 20.9\pm0.2$ ($M_B \le -22.0\pm0.2$; assuming no host extinction),
which is relatively faint. At a similar redshift, only  the afterglow of GRB
050502A was fainter (Kann et al. \cite{Kann2007}).

\subsection{The host galaxy \label{host}}

\begin{figure*}
\centering
\includegraphics[width=13cm,clip=true]{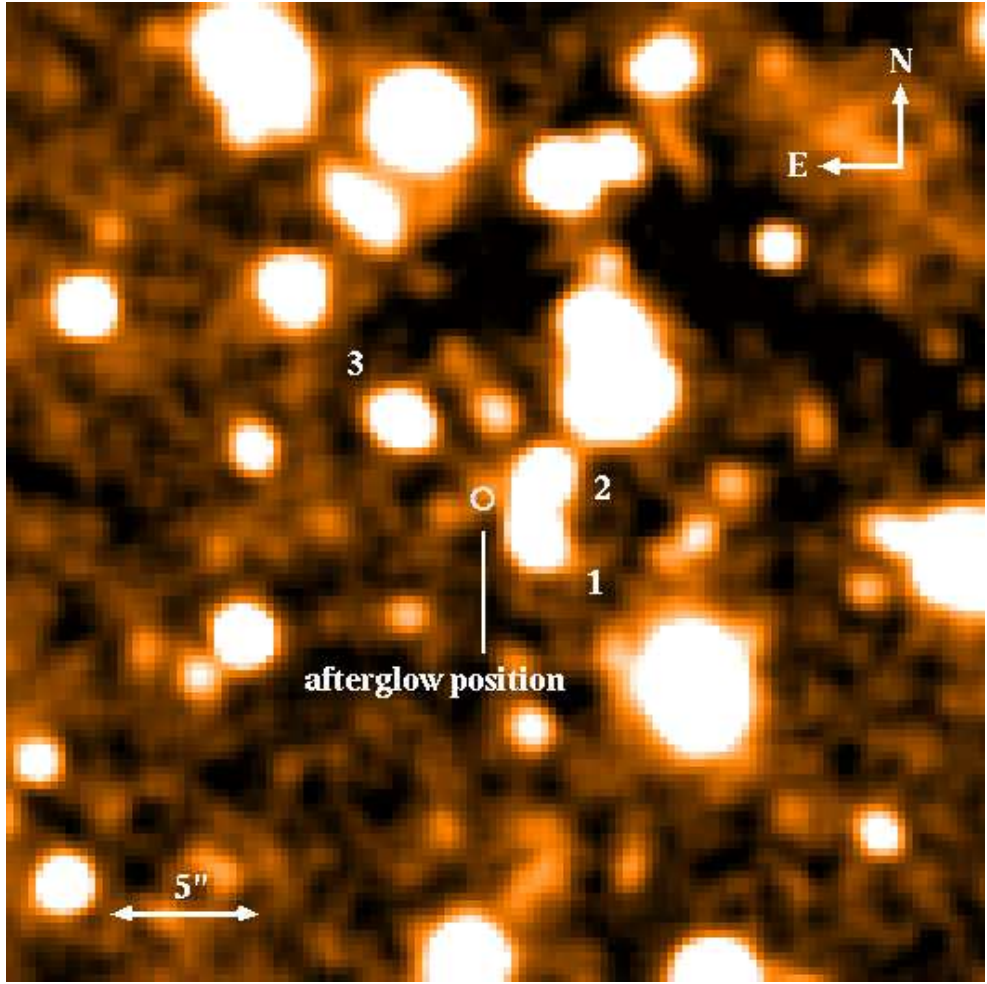}
\caption{
The stellar field around the position of the afterglow of GRB 060605. The
$R_C$-band image is the average of $8\,\times\,500$ s exposure time each,
taken on 14 Sep 2006 and  18 Jun 2007 at VLT/FORS2, when the afterglow had
faded away. The position of the optical transient is indicated. The
faint extended source underlying the position of the optical transient is the
suspected   GRB host galaxy at $z$=3.773.  The three brightest galaxies next to
it are indicated with numbers.}
\label{fors}
\end{figure*}

To search for the host galaxy of GRB 060605 and the potential foreground
absorber detected in our optical spectrum (Fig.~\ref{fig:doublez}) we used
VLT/FORS2 imaging data of the field  obtained under the ESO Large Programme
177.A-0591 (PI: Jens Hjorth). Based on images obtained with the 3.5m Italian
Telescopio Nationale Galileo on La Palma (Malesani et al. 2006; J. Deng et
al., in preparation) we were able to derive an improved astrometric
position  of the optical transient. Its refined position is R.A. (J2000) =
21$^h$ 28$^m$ 37\fs314 and Decl. = --06$^{\circ}$ 3\arcmin 30\farcs88. On the
deep VLT image the afterglow can be positioned with an accuracy of 0.1
arcsec. At this position a very faint extended source is visible
(Fig.~\ref{fors}). Using the average zeropoint of FORS2 $R_C$-band images in
the time period from July to September 2007  ($28.404  \pm 0.037$), as it
is provided on ESO's web pages, we derive a magnitude of $R_C=26.4\pm0.3$
for this source (aperture diameter = 10 pixels). If this is the host galaxy,
it implies that at its peak time the optical afterglow of GRB 060605 was
approximately 11 mag brighter in $R_C$ than its host. We caution, however,
that the detection is weak and we cannot claim that this object is the host of
GRB 0606005 as we have no information about its redshift. On the other hand,
its position underlying the optical transient and its  faint magnitude make a
potential physical association with GRB 060605 a reasonable assumption.

One can also speculate if any of the other three bright, extended sources seen
near the afterglow position on the VLT image could be the host galaxy of GRB
060605. In Fig.~\ref{fors} these three galaxies are indicated with the numbers
1, 2, and 3. However, there are two arguments against this hypothesis. First,
these galaxies have $R_C \sim 24.5$.  For an assumed redshift of $z=3.7$ this
would place all of them at the very bright end of the Schechter luminosity
function (cf. Lin et al. 1996). Second, the angular distance of the  optical
transient from the centres of these galaxies is 2\farcs12, 2\farcs27, and
3\farcs94, respectively. For the considered world model at a redshift of
$z=3.773$  an angular distance of 1 arcsec corresponds to a projected distance
of 7.26 kpc. The projected distance of the  optical transient from the three
galaxies is then 15.4, 16.5 and 28.6 kpc, respectively. Compared to the offset
distribution of GRBs with respect to their host galaxies (in the
pre-\emph{Swift} era; Bloom et al. \cite{Bloom2002}) these large distances
make it unlikely that one of these galaxies is the host. Finally, using
basically the same arguments  it is unlikely that one of them is the
foreground absorber seen in the optical spectrum of the optical transient at
$z$=3.709 (Sect.~\ref{NHoptical}). On the other hand, the foregroud absorber
could be   the faint object to the south-east of the possible host galaxy.

\section{Summary}

We have reduced and analysed XRT and UVOT data from \emph{Swift} and integral
field unit spectra of the afterglow of GRB 060605. In addition, VLT images
were obtained to search for the GRB host galaxy.  We find: (1) The afterglow
spectrum reveals two absorption line systems at redshifts 3.773 and 3.709.  We
identify the former  with the redshift of the burst. (2) The deduced measured
\ion{H}{I} column density for the host galaxy is in between   $N_{\rm HI}$ =
10$^{18.5}$ and 10$^{19.3}$ cm$^{-2}$. It is one of the lowest ever detected
in a GRB afterglow. (3) From the observed time evolution of the X-ray/optical
SED, pointing to a decrease of $\nu_c$ with time, we conclude that the
afterglow propagated into an ISM environment. The cooling frequency was below
the $R$ band  ($\nu_c < \nu_R$) after the jet break time. The initial Lorentz
factor of the fireball was about 250.  (4)  Our analysis of the X-ray light
curve suggests that it followed the canonical X-ray light curve shape (Nousek
et al. \cite{Nousek2006}; Zhang et al. \cite{Zhang2006}). A comparison of the
X-ray and the optical afterglow light curves reveals an  achromatic evolution
at late times.   (5) The observed jet break time is at 0.27 days. This early
jet break time (in the  GRB host frame at about 4900 s after the burst) is the
most remarkable property of GRB 060605. (6) The early observed jet break
translates into a relatively small beaming angle  of 2.4 degrees and hence a
relatively small beaming-corrected energy release in the gamma-ray band of
about 2.1$\,\times\,10^{49}$ erg. These values are not exceptionally low,
however (cf. Zeh et al. 2006; Racusin et al.~\cite{Racusin2008}). In the X-ray
band the afterglow released an energy that is of comparable amount. In the
optical,  at early times, the afterglow was among the most luminous ever
detected. (7) A faint ($R_C=26.4 \pm 0.3$), extended source seen on VLT images
at the position of the afterglow might be the GRB host galaxy or the
foreground absorber seen in the optical spectrum.

The detailed study of this burst was triggered by our goal to use integral
field units to perform rapid follow-up observations of arcsec-sized
\emph{Swift} X-ray error circles.  While  in this particular case our
observations were performed only some hours after the event, we could obtain
useful spectra. One can imagine that a much faster response with an
integral field unit,  immediately after the announcement of an  arcsec-sized
\emph{Swift} X-ray error circle, can provide  early spectral information on
bursts.


\begin{acknowledgements}

P.F., S.K. and D.A.K., acknowledge financial support by DFG grant Kl 766/13-2
and by the German Academic Exchange Service (DAAD) under grant No. D/05/54048.
S.K. and S.S. thank Kim Page (Leicester) for useful discussions.  The research
activities of J.G. are supported  by the Spanish Ministry of Science and
Education through projects AYA2004-01515 and
ESP2005-07714-C03-03. \emph{Swift} is supported at PSU by NASA contract
NAS5-00136. The Dark Cosmology Centre is funded by the Danish National
Research Foundation. We thank the Calar Alto and the ESO staff for excellent
support and the referee for a very careful reading of the manuscript and
very helpful remarks.

\end{acknowledgements}


\end{document}